\documentclass{ptephy}
\usepackage{amsmath,amssymb,bm}

\newcommand{\be}{\begin{equation}}
\newcommand{\ee}{\end{equation}}
\newcommand{\bea}{\begin{eqnarray}}
\newcommand{\eea}{\end{eqnarray}}

\newcommand{\eq}[1]{Eq.~\eqref{eq:#1}}
\newcommand{\sect}[1]{Sec.~\ref{sec:#1}}

\newcommand{\del}{\partial}

\newcommand{\hydroT}{T}
\newcommand{\BYT}{{\mathcal T}}
\newcommand{\BY}{Brown-York }

\newcommand{\Tmn}{\hydroT{}^\mu_{~\nu}}
\newcommand{\Ttt}{\hydroT^t_{~t}}
\newcommand{\Ttz}{\hydroT^z_{~t}}
\newcommand{\Tzz}{\hydroT^z_{~z}}
\newcommand{\Tij}{\hydroT^i_{~j}}
\newcommand{\deltamn}{\delta^\mu_{~\nu}}
\newcommand{\deltaij}{\delta^i_{~j}}

\newcommand{\dTtt}{\delta\hydroT^t_{~t}}
\newcommand{\dTtz}{\delta\hydroT^z_{~t}}
\newcommand{\dTxx}{\delta\hydroT^x_{~x}}
\newcommand{\dTyy}{\delta\hydroT^y_{~y}}
\newcommand{\dTzz}{\delta\hydroT^z_{~z}}
\newcommand{\dTij}{\delta\hydroT^i_{~j}}

\newcommand{\Kmn}{K{}^\mu_{~\nu}}

\newcommand{\BYmn}{\BYT{}^\mu_{~\nu}}
\newcommand{\BYtt}{\BYT{}^t_{~t}}
\newcommand{\BYtz}{\BYT{}^t_{~z}}
\newcommand{\BYij}{\BYT{}^i_{~j}}
\newcommand{\dBYtt}{\delta\BYT{}^t_{~t}}
\newcommand{\dBYtz}{\delta\BYT{}^z_{~t}}
\newcommand{\dBYxx}{\delta\BYT{}^x_{~x}}
\newcommand{\dBYyy}{\delta\BYT{}^y_{~y}}
\newcommand{\dBYzz}{\delta\BYT{}^z_{~z}}
\newcommand{\dBYij}{\delta\BYT{}^i_{~j}}
\newcommand{\dBYmn}{\delta\BYT{}^\mu_{~\nu}}

\newcommand{\tilT}{\tilde{T}}
\newcommand{\tilomega}{\tilde{\omega}}
\newcommand{\tilq}{\tilde{q}}

\newcommand{\tilm}{\tilde{\mu}}
\newcommand{\tili}{\tilde{i}}
\newcommand{\tilt}{\tilde{t}}
\newcommand{\tilz}{\tilde{z}}

\newcommand{\tilhzt}{h^{\tilz}_{~\tilt}}
\newcommand{\tildBYtz}{\delta\BYT{}^{\tilz}_{~\tilt}}

\newcommand{\htt}{h^t_{~t}}
\newcommand{\hxx}{h^x_{~x}}
\newcommand{\hyy}{h^y_{~y}}
\newcommand{\hzz}{h^z_{~z}}
\newcommand{\hzt}{h^z_{~t}}
\newcommand{\hii}{h^i_{~i}}

\newcommand{\hij}{h^i_{~j}}
\newcommand{\hmn}{h^\mu_{~\nu}}
\newcommand{\sumh}{h_s}

\newcommand{\barepsilon}{\bar{\varepsilon}}
\newcommand{\barP}{\bar{P}}
\newcommand{\bars}{\bar{s}}
\newcommand{\barg}{\bar{g}}

\newcommand{\NH}{\text{NH}}

\newcommand{\rc}{r_c}
\newcommand{\uc}{u_c}

\newcommand{\hateta}{\hat{\eta}}
\newcommand{\hatzeta}{\hat{\zeta}}

\newcommand{\nq}{\mathfrak{q}}
\newcommand{\nw}{\mathfrak{w}}

\bmdefine{\bmk}{{\bm{k}}}
\bmdefine{\bmp}{{\bm{p}}}
\bmdefine{\bmq}{{\bm{q}}}
\bmdefine{\bmx}{{\bm{x}}}
\bmdefine{\bmPi}{{\bm{\pi}}}
\bmdefine{\bmo}{{\bm{0}}}
\bmdefine{\bmJ}{{\bm{J}}}
\bmdefine{\bmnabla}{{\bm{\nabla}}}
\bmdefine{\bmf}{{\bm{f}}}
\bmdefine{\bmv}{{\bm{v}}}
\bmdefine{\bmE}{{\bm{E}}}
\begin{document}

\title{The incompressible Rindler fluid versus the Schwarzschild-AdS fluid}


\author{\name{Yoshinori Matsuo}{1}, 
\name{Makoto Natsuume}{1,2,\ast}, 
\name{Masahiro Ohta}{1,2},
and \name{Takashi Okamura}{3},
}

\address{\affil{1}{KEK Theory Center, Institute of Particle and Nuclear Studies, 
High Energy Accelerator Research Organization, 
Tsukuba, Ibaraki, 305-0801, Japan}
\affil{2}{Department of Particle and Nuclear Physics, 
The Graduate University for Advanced Studies, 
Tsukuba, Ibaraki, 305-0801, Japan}
\affil{3}{Department of Physics, 
Kwansei Gakuin University, 
Sanda, Hyogo, 669-1337, Japan}
\email{makoto.natsuume@kek.jp}}

\begin{abstract}
We study the proposal by Bredberg {\it et al.} (1006.1902), where the fluid is defined by the Brown-York tensor on a timelike surface at $r=r_c$ in black hole backgrounds. We consider both Rindler space and the Schwarzschild-AdS (SAdS) black hole. The former describes an incompressible fluid, whereas the latter describes the vanishing bulk viscosity at arbitrary $r_c$. Although the near-horizon limit of the SAdS black hole is Rindler space, these two results do not contradict each other. We also find an interesting ``coincidence" with the black hole membrane paradigm which gives a negative bulk viscosity. In order to show these results, we rewrite the hydrodynamic stress tensor via metric perturbations using the conservation equation. The resulting expressions are suitable to compare with the Brown-York tensor. 
\end{abstract}


\subjectindex{AdS/CFT correspondence, Black holes in string theory} 
\maketitle

\section{Introduction and summary}

According to the AdS/CFT duality \cite{Maldacena:1997re,Witten:1998qj,Witten:1998zw,Gubser:1998bc}, an AdS black hole is dual to a strongly-coupled large-$N_c$ plasma. But as is well-known, it is an old idea that a black hole describes a fluid. There are at least three formulations which realizes this idea:
\begin{enumerate}

\item
Historically, the membrane paradigm \cite{Price:1986yy,Parikh:1997ma} is the oldest one. In this case, the fluid lives on the stretched horizon $r\rightarrow r_0$. However, the membrane paradigm has the unpleasant features to interpret as a fluid such as a {\it negative} bulk viscosity. The membrane paradigm originally focuses on the $(3+1)$-dimensional asymptotically flat black holes, but asymptotics should not matter much since it focuses on the near-horizon limit.
 
\item
In the AdS/CFT duality, the dual fluid ``lives" at the AdS boundary $r\rightarrow\infty$. The advantage of the AdS/CFT duality is a clear microscopic interpretation for the dual fluid. 
The AdS/CFT results are widely used for real-world applications such as the quark-gluon plasma. (See, {\it e.g.}, Refs.~\cite{Natsuume:2007qq,Son:2007vk,CasalderreySolana:2011us} for reviews.) 

\item
More recently, Bredberg, Keeler, Lysov, and Strominger (BKLS) \cite{Bredberg:2010ky,Bredberg:2011jq} proposed the timelike surface at arbitrary position $r=r_c$ for the ``boundary" where the fluid lives (See also, {\it e.g.}, Refs.~\cite{Compere:2011dx,Brattan:2011my}). 
The BKLS approach is analogous to the holographic renormalization. In the near-horizon limit, the BKLS approach describes an {\it incompressible} fluid. 

\end{enumerate}
Another closely related idea is a ``black hole in a cavity" \cite{York:1986it}. This idea was proposed to obtain a well-defined thermal equilibrium for asymptotically flat black holes such as the Schwarzschild black hole. The Schwarzschild black hole has a negative heat capacity, so it is unstable by Hawking radiation. However, if the black hole is surrounded by a finite-temperature cavity, and if the cavity is close enough to the horizon, a thermal equilibrium is achieved. In a sense, the BKLS approach is an AdS black hole in a cavity. 

While each approach has a different motivation and physical interpretation, they have one thing in common: they all employ the \BY tensor \cite{Brown:1992br} as the fluid stress tensor. Thus, they are somehow related to each other. 

Both in the membrane paradigm and in the BKLS approach (in particular in Ref.~\cite{Bredberg:2011jq}), one often starts to identify the velocity field of the fluid in the bulk spacetime. This has its own advantage that the relation between the Einstein equation and the Navier-Stokes equation is direct and transparent.
On the other hand, this brings us to an immediate problem of why a particular vector field should be regarded as the velocity field. So, we do not take such a path.
\begin{itemize}

\item
Instead, we consider metric perturbations and study the (linear) response of the \BY tensor by the perturbations {\it \`{a} la} AdS/CFT duality. 

\item
In hydrodynamics, the velocity field is determined from the metric perturbations (\sect{hydro}). Then, one can eliminate the velocity field completely in the hydrodynamic stress tensor. The resulting expression contains metric perturbations only, which is suitable to compare with the \BY tensor. In our approach, the velocity field is a consequence of metric perturbations. 

\end{itemize}
One purpose of this paper is to reexamine the BKLS approach using the above formulation.

In particular, we study the issue of the bulk viscosity $\zeta$, which is non-negative in the AdS/CFT duality, negative in the membrane paradigm, and is irrelevant in the BKLS approach (because of an incompressible fluid). For that purpose, we consider the sound mode perturbations whose analysis was somewhat incomplete in Ref.~\cite{Bredberg:2010ky}. 

We study Rindler space, which is the near-horizon limit of black holes with nondegenerate horizon, and the five-dimensional Schwarzschild-AdS black hole (SAdS$_5$)%
\footnote{While our work was in progress, there appeared preprints which study Rindler hydrodynamics \cite{Marolf:2012dr,Eling:2012ni,Eling:2012xa}. }.
The near-horizon limit of the SAdS$_5$ black hole is Rindler space (\sect{comparison}). So, one expects that the bulk viscosity for the SAdS$_5$ black hole agrees with the Rindler result in the limit $r_c\rightarrow r_0$. In the AdS/CFT duality, the bulk viscosity for the SAdS$_5$ black hole vanishes in the limit $r_c\rightarrow\infty$ because of the scale invariance of the geometry. However, when $r_c\neq\infty$, the stress tensor for the SAdS$_5$ black hole is no longer traceless [\eq{SAdS_thermodynamic}], so one must examine the bulk viscosity in this case. 
Our results are summarized as follows:
\begin{enumerate}

\item
For Rindler space, the \BY tensor gives an incompressible fluid in accordance with the BKLS result (\sect{Rindler}).
 
\item
For the SAdS$_5$ black hole, the \BY tensor always gives the vanishing bulk viscosity irrespective of the boundary position $r_c$ (\sect{SAdS5}).

\item
There are no contradictions between two results since the hydrodynamic regime used for the SAdS black hole ``differs" from the hydrodynamic regime used for Rindler space (when expressed in terms of the SAdS variables) (\sect{comparison}). 

\end{enumerate}
We also find an interesting ``coincidence" with the membrane paradigm in the Rindler analysis. If one does not take into account a constraint equation of the Einstein equation (in hydrodynamics, this corresponds to not taking the continuity equation into account), one would get the negative bulk viscosity in accordance with the membrane paradigm (\sect{constraints}). The precise relation to the membrane paradigm is not clear and is left to a future work.  
In addition, we obtain one of the second-order hydrodynamic transport coefficients $\tau_\pi$ for the SAdS$_5$ black hole in the BKLS approach.

\section{Linearized hydrodynamics by metric perturbations}\label{sec:hydro}

\subsection{Homogeneous perturbations}

The basic hydrodynamic equation is the conservation equation 
\be
\nabla_\mu \hydroT^{\mu\nu}=0
%
\ee
(or the continuity equation and the Navier-Stokes equation). In $(3+1)$-dimensions, there are 4 equations whereas the stress tensor has 10 components. Since the equation of motion is not closed, one introduces the constitutive equation%
\footnote{We use $\mu, \nu, \ldots$ for the $(p+1)$-dimensional boundary coordinate indices. The boundary spatial coordinates $x_i$ are also denoted as $x_i=(x,y,z)$ for $p=3$. We use indices $a,b, \ldots$ for the spatial coordinates transverse to $z$. }:
\begin{align}
\hydroT^{\mu\nu} &= (\varepsilon+P)u^\mu u^\nu + P g^{\mu\nu} +\tau^{\mu\nu}~,
\label{eq:hydro_curved} \\
\tau^{\mu\nu} &:= 
- P^{\mu\alpha} P^{\nu\beta} \left[
\eta \left( \nabla_\alpha u_\beta \!+\! \nabla_\beta u_\alpha 
- \frac{2}{p} g_{\alpha\beta}  \nabla_\lambda u^\lambda \right) 
+ \zeta g_{\alpha\beta} \nabla_\lambda u^\lambda 
\right]~,
\end{align}
where 
$P^{\mu\nu} := g^{\mu\nu} + u^\mu u^\nu$ is the projection tensor, $\eta$ is the shear viscosity, and $\zeta$ is the bulk viscosity. 
In other words, one chooses the velocity field $u^\mu$ and the pressure $P$ as the basic hydrodynamic variables. Note that $\varepsilon$ and $P$ are not independent; rather they are determined by an equation of state. We choose $P$ as the independent variable. We assume $\varepsilon=\varepsilon(P)$ and use $c_s^2=\del P/\del\varepsilon$, where $c_s$ is the speed of sound. Then, there are 4 degrees of freedom in $(3+1)$-dimensions (three from $u^\mu$ because of $u^2=-1$ and one from $P$), and the equation of motion is closed.

In equilibrium, there is no spatial flow, so one can take the rest frame $u^i=0$. Then, one has
\be
\Ttt = - \barepsilon~, \qquad \Tij = \barP \deltaij~, 
%
\ee
where `` $\bar{~}$ " denotes an equilibrium value%
\footnote{In this paper, we consider $\Tmn$, the stress tensor with one upper and one lower indices, which is convenient to compare with the \BY tensor (\sect{Rindler_thermo}).}.

When one adds external gravitational perturbations $h^\mu_{~\nu}$, the hydrodynamic variables $P$ and $u^i$ have responses following the conservation equation. By solving the conservation equation, one can determine the responses. For hydrodynamic computations, we always use the Minkowski background $\barg_{\mu\nu}=\eta_{\mu\nu}$. We consider metric perturbations of the form 
\begin{equation}
\hmn(t,z) = \hmn e^{-i\omega t+iqz}~.
\end{equation}
Then, the metric perturbations are decomposed as the tensor, shear and sound modes. We consider the sound mode, which consists of
\be
\htt~, \qquad
h^a_{~a} = \hxx~, \qquad
\hzz~, \qquad
\hzt
\ee
(no summation over the index $a$).
The metric becomes
\be
ds^2 = - (1 + \htt) dt^2 + \sum_i (1+\hii) dx_i^2 + 2\hzt dt dz~.
%
\ee
We write the responses as
\be
P(t,z) = \barP + \delta P e^{-i\omega t+iqz}~, \qquad
u^i (t,z) = \delta u^i e^{-i\omega t+iqz}~, 
%
\ee
and $\varepsilon (t,z) = \barepsilon + \delta\varepsilon e^{-i\omega t+iqz}$. Accordingly, the stress tensor has the response
\be
\Tmn(t,z) = \bar{\hydroT}^\mu_{~\nu} + \delta\Tmn e^{-i\omega t+iqz}~.
%
\ee

We first consider homogeneous perturbations $q=0$. Since $u^2=-1$, $u^t = 1-\htt/2$ (one can set $u^a=0$). From the conservation equation $\nabla_\mu \hydroT^{\mu\nu}=0$, one gets
\begin{align}
& i\omega \left\{ \delta\varepsilon+ \frac{\barepsilon+\barP}{2} \sumh \right\} = 0~, 
\label{eq:continuity} \\
& i\omega (\barepsilon+\barP) (\hzt+\delta u^z) = 0~,
%
\end{align}
where $\sumh := \sum_k h^k_{~k}$ is the {\it spatial} trace. Then, $\delta\Tmn$ becomes
\begin{subequations}
\label{eq:hydro_q=0}
\begin{align}
\dTtt &= -\delta\varepsilon = \frac{\barepsilon+\barP}{2} \sumh~, 
\label{eq:hydro_00} \\
\dTtz &= (\barepsilon+\barP) \hzt~,
\label{eq:hydro_0z} \\
\dTij &= \delta P(h) \deltaij + i\eta\omega \hij 
 - i\left(\frac{\eta}{p} - \frac{\zeta}{2} \right)\omega \sumh \deltaij~,
\label{eq:hydro_ij} 
\end{align}
\end{subequations}
where
\be
\delta P(h) = c_s^2 \delta\varepsilon = -\frac{\barepsilon+\barP}{2} c_s^2 \sumh~.
%
\ee
These expressions may be familiar to readers. For instance, see App.~A of Ref.~\cite{Gubser:2008sz} for $\dTij$. However, the inhomogeneous perturbation case ($q\neq0$) in the next subsection is more involved and deserves close inspection. 

Anticipating the bulk results in the following sections, let us consider the $c_s\rightarrow\infty$ limit. 
In the $c_s\rightarrow\infty$ limit, the continuity equation \eqref{eq:continuity} becomes 
\be
i\omega \barP \sumh = 0~.
%
\ee
The $c_s\rightarrow\infty$ limit is rather special. 
In this limit, the conservation equation gives a condition for the perturbations instead of a response.  
In order that time-dependent perturbations are allowed, the spatial perturbations must be traceless. Or the fluid must be compressible for generic time-dependent homogeneous perturbations. 
Then, one obtains
\begin{subequations}
\label{eq:hydro_incompressible}
\begin{align}
\dTtt &\rightarrow 0~, 
\\
\dTtz &\rightarrow (\barepsilon+\barP) \hzt~,
\\
\dTij &\rightarrow i\eta\omega \hij~.
%
\end{align}
\end{subequations}

\subsection{Inhomogeneous perturbations}

We turn to inhomogeneous perturbations $q\neq0$. 
Again take $u^t=1-\htt/2$.
For $q\neq0$, the continuity equation becomes
\be
- i\omega \left(\frac{\delta\varepsilon}{\barepsilon+\barP} + \frac{1}{2} \sumh \right) + iq \delta u^z = 0~.
%
\ee
Combining this with the Navier-Stokes equation gives
\begin{subequations}
\label{eq:response_q}
\begin{align}
\delta u^z &= \frac{\omega}{q} 
  \frac{c_s^2 q^2}{c_s^2 q^2 - \omega^2 - i\Gamma_s\omega q^2}
  \left[ \frac{1}{2} \sumh + \frac{\omega}{c_s^2 q} \hzt - \frac{\htt}{2c_s^2} \right.
\nonumber \\
&\quad
 \left.
  - \frac{i}{c_s^2} \left\{ \frac{1}{2}\left(\hatzeta -\frac{2}{3} \hateta\right)\omega (\hxx+\hyy) 
  + \frac{\Gamma_s}{2} \omega \hzz
  \right \} \right]~,
\label{eq:response_q1}
\\
\delta P &= (\barepsilon+\barP)
  \frac{c_s^2 q^2}{c_s^2 q^2 - \omega^2 - i\Gamma_s\omega q^2}
  \left[ - \frac{1}{2} \htt + \frac{\omega}{q} \hzt 
  + \frac{\omega^2}{2q^2} \sumh
  + i \hateta \omega \left( \hxx+\hyy \right) \right]~,
\label{eq:response_q2}
\\
\hateta &:= \frac{\eta}{\barepsilon+\barP}~, \quad 
\hatzeta := \frac{\zeta}{\barepsilon+\barP}~, \quad
\Gamma_s := \frac{1}{\barepsilon+\barP}\left(\frac{4}{3}\eta+\zeta\right)~,
%
\end{align}
\end{subequations}
where $\Gamma_s$ is the sound attenuation constant.  Also, these are momentum-space expressions so are complex; in real-space, they are real.

Having written down all hydrodynamic variables via metric perturbations, we are ready to express the hydrodynamic stress tensor via metric perturbations only. The full expression is rather cumbersome, so we give the expressions only in the $c_s\rightarrow\infty$ limit, which is relevant to the Rindler case.
In the $c_s\rightarrow\infty$ limit, 
\begin{align}
\delta u^z &\rightarrow \frac{\omega}{2q} \sumh~,
\label{eq:response_q_incompressible1}
\\
\delta P &\rightarrow (\barepsilon+\barP)
  \left[ - \frac{1}{2} \htt + \frac{\omega}{q} \hzt + \frac{\omega^2}{2q^2} \sumh \right]
  + i \eta\omega \left( \hxx+\hyy \right)~.
\label{eq:response_q_incompressible2}
\end{align}
Note that the sound pole $(c_s^2 q^2 - \omega^2 + i\Gamma_s\omega q^2)^{-1}$ in Eqs.~\eqref{eq:response_q} disappears in this limit. At the same time, the dependence on the bulk viscosity disappears. 
As a check, substituting \eq{response_q_incompressible1} into the continuity equation gives $\delta\varepsilon=0$ as expected. Also, some components of covariant derivatives are
\begin{align}
\nabla_x u^x &= -\frac{1}{2} i \omega \hxx~, &
\nabla_y u^y &= -\frac{1}{2} i \omega \hyy~, &
\nabla_z u^z &= \frac{1}{2} i \omega (\hxx+\hyy)~, &
%
\end{align}
so $u^\mu$ obeys 
\begin{equation}
 \nabla_\mu u^\mu = 0~.
 \label{eq:incompressible}
\end{equation}
Then, $\delta\hydroT^\mu_{~~\nu}$ becomes
\begin{subequations}
\label{eq:hydro_q_incompressible}
\begin{align}
\dTtt &\rightarrow 0~, 
\\
\dTtz &\rightarrow -(\barepsilon+\barP) \frac{\omega}{2q} \sumh~, 
\\
\dTxx &\rightarrow \delta P(h) + i\eta\omega \hxx~,
\\
\dTyy &\rightarrow \delta P(h) + i\eta\omega \hyy~,
\\
\dTzz &\rightarrow \delta P(h) - i\eta\omega (\hxx+\hyy)~,
%
\end{align}
\end{subequations}
in the $c_s\rightarrow\infty$ limit. Several points of Eqs.~\eqref{eq:hydro_q_incompressible} deserve comment. (i) $\Ttz$ is not proportional to $\hzt$ [{\it cf.}, \eq{hydro_0z}]. (ii) The $O(i\omega)$ terms  of $\Tzz$ are not proportional to $\hzz$. (iii) While Eqs.~\eqref{eq:response_q} themselves have a well-defined $q\rightarrow0$ limit, the $c_s\rightarrow\infty$ case does not have a limit. So, we consider the $q=0$ and $q\neq0$ cases separately. 

We now compare hydrodynamic expressions obtained in this section with the \BY tensor in Rindler space and in the SAdS$_5$ black hole.

\section{Sound mode in Rindler space}\label{sec:Rindler}

\subsection{Thermodynamic quantities}\label{sec:Rindler_thermo}

The $(p+2)$-dimensional Rindler space is given by 
\begin{equation}
 ds_{p+2}^2 
  = 
  -r dt^2 + \frac{dr^2}{r} 
  + \sum_i d x_i^2~. 
\end{equation}
The Rindler horizon is located at $r=0$ and the Hawking temperature is given by $T=1/(4\pi)$.

We consider the timelike surface $r=r_c$. The \BY tensor is given by
\begin{equation}
\BYmn= \frac{1}{8\pi G}\left(\deltamn K - \Kmn \right)~,
\label{eq:BYtensor}
\end{equation}
where $\Kmn$ is the extrinsic curvature of the surface, and $K$ is its trace. In this paper, we denote the \BY  tensor as $\BYmn$ to avoid confusion with the hydrodynamic stress tensor $\Tmn$. For a diagonal metric, the extrinsic curvature takes a simple form:
\begin{equation}
\Kmn = \frac{1}{2} n^r g^{\mu\rho} \partial_r g_{\rho\nu}~,
\label{eq:extrinsic_diagonal}
\end{equation}
where $g_{\mu\nu}$ is the induced metric on the surface. For Rindler space, $g_{\mu\nu} = \text{diag}(-\rc, \bm{1})$. Also, $n^r$ is the unit normal to the $r=r_c$ surface: $n^r = 1/\sqrt{g_{rr}}$. 

We consider the \BY tensor with one upper and one lower indices from the following reasons:
\begin{enumerate}
\item
The counterterm takes the form ${\BYmn}^{\text{(CT)}} \propto \delta^\mu_{~\nu}$ (see below), so the counter term dependence is absent upon metric perturbations.
\item
We chose the Minkowski background $\barg_{\mu\nu} = \eta_{\mu\nu}$ for the hydrodynamic computations. But this differs from the induced metric used for the \BY tensor by $r$-rescaling, {\it e.g.}, $g_{\mu\nu} = \text{diag}(-\rc, \bm{1})$ for Rindler space. One way is to transform the \BY tensor from the original coordinates $x^\mu$ to the proper coordinates $x^{\tilm}$:
\be
\tilt =\sqrt{-\barg_{tt}}\, t~, \qquad x^{\tili} = \sqrt{\barg_{ii}}\, x^i~.
%
\ee
However, it is not necessary to distinguish $x^\mu$ and $x^{\tilm}$ for $\BYtt$ and $\BYij$ since the upper and lower indices receive the opposite scaling. (This does not apply to $\BYtz$, so care is necessary.)
\end{enumerate}

One can add ``counterterms" to the \BY tensor.
From the AdS/CFT point of view, the counterterms regularize divergences in physical quantities \cite{Balasubramanian:1999re}. They are given by
\begin{equation}
{\BYmn}^{\text{(CT)}} =  -\frac{1}{16\pi G}
\left(c_1 \deltamn +c_2 {G^{\mu}_{~\nu}}^{(p+1)} + \cdots \right)~.
\label{eq:CT}
\end{equation}
The coefficient $c_1=2p/L$, where $L$ is the AdS radius. 
${G^{\mu}_{~\nu}}^{(p+1)}$ is the Einstein tensor built from the induced metric $g_{\mu\nu}$. 
For Rindler space, ${G^{\mu}_{~\nu}}^{(p+1)}$ vanishes since the surface is flat%
\footnote{This will not be the case when one adds metric perturbations. But our primary concern is thermodynamic quantities and transport coefficients. The transport coefficients of first-order hydrodynamics appear only in $O(\omega)$ terms in the stress tensor, while $G_{\mu\nu}^{(p+1)}$ gives  $O(\omega^2, q^2)$ terms, so we can safely ignore the Einstein tensor. }.
We include the boundary cosmological constant term in order to compare with the SAdS$_5$ result (\sect{SAdS5}).

For Rindler space, $\Kmn=\text{diag} (1/(2\rc^{1/2}), \bm{0})$. Then, one gets
\be
\BYtt = -\frac{c_1}{16\pi G}~, \qquad
\BYij = \frac{1}{16\pi G}\left(\frac{1}{\sqrt{\rc}} -c_1 \right) \deltaij~,
%
\ee
which gives
\be
\barepsilon = \frac{c_1}{16\pi G}~, \qquad
\barP = \frac{1}{16\pi G}\left(\frac{1}{\sqrt{\rc}} -c_1 \right) = \frac{\tilT}{4 G} - \frac{c_1}{16\pi G}~.
\label{eq:Rindler_thermodynamic}
\ee
Here, $\tilT$ is the proper temperature not the Hawking temperature $T$:
\be
\tilT (\rc) = \frac{1}{\sqrt{-\barg_{tt}(\rc)}} T~.
%
\ee
When $c_1=0$, $\barP$ agrees with the membrane paradigm result \cite{Price:1986yy,Parikh:1997ma}. 
The thermodynamic relation $\tilT\bars = \barepsilon+\barP$ gives the entropy density $\bars = 1/(4G)$. Since the energy density is constant, the stress tensor describes an incompressible fluid, and the speed of sound $c_s^2 = \partial \barP/\partial \barepsilon$ diverges. 

\subsection{Sound mode perturbations}\label{sec:Rindler_q=0}

We consider sound mode perturbations in Rindler space. We take the gauge where
$ h{}_{* r} = 0 $
for all $_{*}$, and the metric is given by
\be
ds_{p+2}^2 = - r(1 + \htt) dt^2 + \sum_i (1+\hii) dx_i^2 + 2\hzt dt dz + \frac{dr^2}{r}~.
%
\ee
We consider perturbations of the form 
\begin{equation}
\hmn(t,z,r) = \hmn(r)\, e^{-i\omega t+iqz}~.
\end{equation}

In the gauge $h_{*r} = 0$, the extrinsic curvature takes the simple form \eqref{eq:extrinsic_diagonal}. The response of the \BY tensor is given by  ($'=\del_r$)
\begin{subequations}
\label{eq:BY_Rindler}
\begin{align}
\dBYtt &= (\barepsilon+\barP) \rc {\sumh}'~, 
  \label{eq:BY_Rindler_00} \\
\dBYtz &= (\barepsilon+\barP)(\hzt -\rc{\hzt}')~, 
  \label{eq:BY_Rindler_0z} \\
 \dBYij &= 
  (\barepsilon+\barP) \rc \left[   
  - {\hij}' + \deltaij ({\htt}'+{\sumh}')
   \right]~,
  \label{eq:BY_Rindler_ij}
\end{align}
\end{subequations}
where we used $\barepsilon+\barP = 1/(16\pi G \rc^{1/2})$. In order to compare the \BY tensor with the hydrodynamic tensor, one needs to rewrite ${\hmn}'$. This requires the Einstein equation.

We first consider homogeneous perturbations $q=0$. The Einstein equation with $q=0$ gives
\begin{subequations}
\begin{align}
{\hii}'' + \frac{1}{r}{\hii}' + \frac{\omega^2}{r^2} \hii &= 0~. 
\label{eq:master4q=0} \\
(r^{3/2} {\htt}')' &= 0~,
\label{eq:htt4q=0} \\
- 2 \omega {\hzt}' &= 0~,
\label{eq:Ezr4q=0} \\
r {\sumh}' + O(\omega^2) &= 0~.
\label{eq:Err4q=0} 
\end{align}
\end{subequations}
Equations~\eqref{eq:Ezr4q=0} and \eqref{eq:Err4q=0} are two of the ``constraint equations" which are first-order differential equations%
\footnote{In this paper, we use the word ``constraint equations" in the sense of the radial foliation, not the time foliation.}.
The solution of \eq{master4q=0} is given by
\begin{equation}
\hii(r) = \hii(r_c) \left(\frac{r}{r_c}\right)^{-i\omega}~,
\label{eq:Sol4q=0}
\end{equation}
where we imposed the ``incoming wave" boundary condition at the horizon. The remaining integration constant is fixed by the Dirichlet boundary condition $\hii(r_c)$. Equation~\eqref{eq:Sol4q=0} is the exact solution for all $r$. For $\htt$, imposing the regularity condition at the horizon, one gets $\htt = \htt(r_c)$. 

From \eq{Err4q=0}, we obtain $\dBYtt=0$, so the \BY tensor describes an incompressible fluid. From \eq{Err4q=0} and ${\htt}'=0$, the terms proportional to $\deltaij$ vanish, which implies an incompressible fluid as well. 
Finally, from \eq{Ezr4q=0}, a non hydrodynamic term in $\dBYtz$ [the second term of \eq{BY_Rindler_0z}] vanishes.

Thus, the \BY tensor becomes
\begin{subequations}
\label{eq:BY_Rindler_q=0}
\begin{align}
\dBYtt &= 0~, 
\\
\dBYtz &= (\barepsilon+\barP)\hzt~, 
\\
 \dBYij &= \frac{i\tilomega}{16\pi G} \hij~,
%
\end{align}
\end{subequations}
where we used \eq{Sol4q=0} and $\tilomega$ is the proper frequency. 
In order to compare the \BY tensor with the Minkowski hydrodynamic stress tensor \eqref{eq:hydro_q=0}, one needs to rewrite the \BY tensor in proper coordinates $\tilt = \sqrt{-\barg_{tt}}\, t$ and $x^{\tili}=\sqrt{\barg_{zz}}\, x^i$. In this paper, ``~$\tilde{~}$~" denotes proper coordinates and proper quantities. Proper frequencies and wave numbers are given by
\be
\tilomega = \frac{\omega}{\sqrt{-\barg_{tt}}} = \frac{\omega}{\rc^{1/2}}~, \qquad
\tilq = \frac{q}{\sqrt{\barg_{zz}}} = q~.
\label{eq:proper}
\ee
However, as discussed previously, it is not necessary to distinguish $x^\mu$ and $x^{\tilm}$ for $\dBYtt$ and $\dBYij$ except the replacement \eqref{eq:proper}. For the off-diagonal component, $\dBYtz \propto \hzt$, so the expression does not change under the coordinate transformation. 

Equations~\eqref{eq:BY_Rindler_q=0} take the same form as the hydrodynamic stress tensor in the $c_s\rightarrow\infty$ limit \eqref{eq:hydro_q_incompressible} with 
\be
\eta = \frac{1}{16\pi G}~.
%
\ee
This agrees with the membrane paradigm result and the BKLS result \cite{Price:1986yy,Parikh:1997ma,Bredberg:2010ky,Bredberg:2011jq}. On the other hand, the result of an incompressible fluid differs from the membrane paradigm.  

From Eqs.~\eqref{eq:BY_Rindler}, the term ${\sumh}'$ gives $\delta\varepsilon$ and the bulk viscosity, but ${\sumh}'=0$ up to first order in $(\omega, q)$, so one immediately has an incompressible fluid. This is true even for $q\neq0$ [\eq{Err}], thus one expects that the fluid remains incompressible even for $q\neq0$. However, it is not obvious that the \BY tensor takes the same form as the hydrodynamic tensor when $q\neq0$. Thus, we turn to the $q\neq0$ case in \sect{Rindler_q}. 

\subsection{Possible connection with the membrane paradigm?}\label{sec:constraints}

Our result shows that the \BY tensor gives an incompressible fluid, which differs from the membrane paradigm. However, there is an interesting ``coincidence" with the membrane paradigm if one ignores part of the Einstein equation. 

Let us ignore the constraint equation \eqref{eq:Err4q=0} for a moment, which gives the incompressible condition. In hydrodynamic analysis, the incompressible condition comes from the continuity equation (\sect{hydro}), so ignoring the constraint equation \eqref{eq:Err4q=0} corresponds to ignoring the continuity equation. From \eq{Sol4q=0}, ${\sumh}' = -i\omega \sumh/r$. Substituting this into \eq{BY_Rindler_ij} gives
\be
\dBYij = \frac{i\tilomega}{16\pi G} \left[ \hij - \deltaij \sumh \right]~.
\label{eq:w/o_constraint}
\ee
If one compares this with the hydrodynamic stress tensor \eqref{eq:hydro_q=0}, one would get
\be
\zeta = -\frac{p-1}{p}\frac{1}{8\pi G}~,
\label{eq:membrane_zeta}
\ee
which coincides with the membrane paradigm \cite{Price:1986yy,Parikh:1997ma,Eling:2009sj}. The original membrane paradigm focuses on the $(3+1)$-dimensional case, but the extension into the generic dimensions exists \cite{Eling:2009sj}. Note that $\zeta<0$. 

This is an interesting coincidence, and the result may have some relevance with the membrane paradigm. On the other hand, we should stress that this result itself does not give a consistent hydrodynamic interpretation completely. For example, \eq{w/o_constraint} seems to lack the $\delta P$ term in \eq{hydro_ij}. Also, $\dBYtt$ is nonvanishing, but
\be
\dBYtt = -(\barepsilon+\barP) i\omega\sumh~.
%
\ee
Comparing this with Eqs.~\eqref{eq:hydro_q=0}, this is consistent only if $i\omega=-1/2$. But this brings us another issue. First, we consider the hydrodynamic limit $|\omega|\rightarrow0$, so it is not clear if such an interpretation is possible. Second,  when $|\omega|$ is not small, it is not clear if the $O(\sumh)$ term in \eq{w/o_constraint} is really the viscosity term: the first term and the third term of \eq{hydro_ij} are not distinguishable. 

Thus, the only consistent interpretation is the incompressible fluid by taking \eq{Err4q=0} into account. But the coincidence \eqref{eq:membrane_zeta} is suggestive. This might indicate that the membrane paradigm is not fully consistent. 

\subsection{Inhomogeneous perturbations}\label{sec:Rindler_q}

The Einstein equation consists of second-order differential equations which are dynamical equations and first-order differential equations which are constraint equations. 
The dynamics of the field obeying the constraints is determined by one dynamical equation. They are referred as the master field and the master equation, respectively. This counting goes as follows:
\begin{itemize}

\item
For the sound mode in 5-dimensional spacetime, 4 components of metric perturbations are relevant. 
\item
The Einstein equation gives 4 dynamical equations and 3 constraint equations. Thus, one obtains 1 master equation, which gives the solution for a combination of 4 metric components. 
\item
The solution of the master equation has two integration constants. One is fixed by imposing the incoming wave boundary condition at the horizon. Thus, one obtains the solution for a combination of 4 metric components with one integration constant. 
\item
The other 3 components are calculated using 3 constraint equations which give one integration constant for each component.
\item 
In summary, we obtain 4 solutions with 4 integration constants. These integration constants are fixed by imposing boundary conditions at the boundary on each components, $\hmn(r_c)$. 

\end{itemize}
In reality, in order to compute the \BY tensor, one only needs ${\hmn}'(r_c)$. They can be determined from the constraint equations and the master equation. So, one does not have to solve the constraint equations.

In Rindler space, the constraint equations are given by%
\footnote{These equations correspond to $(t,r)$, $(z,r)$, and $(r,r)$-components of the Einstein equation, respectively.}
\begin{subequations}
\label{eq:constraints}
\begin{align}
\omega \left(2r {\sumh}' - \sumh \right) + 2q \left( r {\hzt}' - \hzt \right) &= 0~,
\label{eq:Etr} \\
- 2 \omega {\hzt}' + q \left( 2r {\htt}'+ \htt + 4r {\hxx}' \right) &= 0~,
\label{eq:Ezr} \\
r {\sumh}' + O(\omega^2, \omega q, q^2) &= 0~.
\label{eq:Err} 
\end{align}
\end{subequations}
The master field is $\hxx$ which obeys
\begin{equation}
{h{}^x{}_x}'' + \frac{1}{r}{h{}^x{}_x}' + \frac{\omega^2 - r q^2}{r^2} h{}^x{}_x = 0~. 
\label{eq:master}
\end{equation}
We solve the master equation by imposing (i) the incoming wave boundary condition at the horizon and (ii) the Dirichlet boundary condition at $r=r_c$, $h{}^x{}_x(r_c)$. After imposing the former boundary condition, the solution takes the form
\begin{align}
\hxx(r) &= \left.\frac{\hxx}{F}\right|_{r_c} F(r)~, 
\label{eq:BC_Rindler}
\end{align}
where we fixed the remaining overall integration constant by the Dirichlet boundary condition $\hxx(r_c)$. The solution of the master equation in the near-horizon limit $r\rightarrow0$ is given by \eq{Sol4q=0}. When $q=0$, it is the exact solution for all $r$. This is not the case when $q \neq 0$, but the master equation takes the form
\be
{\hxx}'' + \frac{1}{r} {\hxx}' + O(\omega^2, q^2) = 0~.
%
\ee
Thus, \eq{Sol4q=0} still gives the solution to first order in $(\omega, q)$:
\begin{equation}
F(r) = 1 - i \omega \log r + O(\omega^2,q^2)~. 
\label{eq:master_sol_q}
\end{equation}

For the sound mode, $h{}^a{}_a = \hxx$, but it is convenient to keep each components separately. 
Also, we focus on the five-dimensional case ($p=3$) for simplicity. Substitute Eqs.~\eqref{eq:constraints} and \eqref{eq:BC_Rindler} into Eqs.~\eqref{eq:BY_Rindler}. To first order in $(\omega, q)$, the  \BY tensor becomes
\begin{subequations} 
\label{eq:BY_Rindler_q}
\begin{align}
\dBYtt &= 0~, 
 \\
\tildBYtz &= \frac{1}{\rc^{1/2}} \dBYtz = - (\barepsilon+\barP) \frac{\tilomega}{2\tilq} \sumh~, 
\\
\dBYxx &= \delta P(h) + \frac{i\tilomega} {16\pi G} \hxx~,
 \\
 \dBYyy &= \delta P(h) + \frac{i\tilomega }{16\pi G} \hyy~,
 \\
 \dBYzz &= \delta P(h) - \frac{i\tilomega }{16\pi G} \left(\hxx + \hyy\right)~,
\end{align}
\end{subequations}
where 
\begin{align}
\delta  P &=  (\barepsilon+\barP) \left[
  - \frac{1}{2} \htt
  + \frac{\omega}{q\rc} \hzt 
  + \frac{\omega^2}{2q^2\rc} \sumh
\right]
  + \frac{i\omega}{16\pi G \rc^{1/2}} \left( \hxx+\hyy \right)~, \\
& =  (\barepsilon+\barP) \left[
  - \frac{1}{2} \htt
  + \frac{\tilomega}{\tilq} \tilhzt 
  + \frac{\tilomega^2}{2\tilq^2} \sumh
\right]
  + \frac{i\tilomega}{16\pi G} \left( \hxx+\hyy \right)~.
%
\end{align}
Again, the \BY tensor in proper coordinates $x^{\tilm}$ takes the same form except for $\dBYtz$. So, we have rewritten $\dBYtz$ (and $\delta P$)  in proper coordinates. The \BY tensor takes the same form as the hydrodynamic stress tensor in the $c_s\rightarrow\infty$ limit \eqref{eq:hydro_q_incompressible} with $\eta=1/(16\pi G)$.

\section{Sound mode in Schwarzschild-AdS black hole}\label{sec:SAdS5}



\subsection{Thermodynamic quantities}

The SAdS$_5$ metric is given by 
\begin{align}
ds_5^2 &= \left(\frac{r}{L}\right)^2 [-f(r) dt^2 + dx_i^2]
+ \frac{dr^2}{ \left(\frac{r}{L}\right)^2 f(r) }~, &
f(r) &= 1 - \left( \frac{r_0}{r} \right)^4~, \\
&= \frac{1}{u}\left[-f(u) dt^2 + dx_i^2\right] + \frac{du^2}{4u^2f(u)}~, &
f(u) &= 1 - u^2~,
%
\end{align}
where $u=(r_0/r)^2$. The Hawking temperature is given by $T=r_0/(\pi L^2)$. We take the horizon radius $r_0=1$ by rescaling $t$ and $x_i$, and we set the AdS radius $L=1$. The boundary position will be denoted as $u = u{}_c$. 

The \BY tensor and thermodynamic relations give the following thermodynamic quantities:
\begin{subequations}
\label{eq:SAdS_thermodynamic}
\begin{align}
 \tilT &= \frac{1}{\pi}\sqrt{\frac{u_c}{1-u_c^2}}~, \\ 
 \barepsilon &= \frac{3}{8\pi G}
 \left( \frac{c{}_1}{6} - \sqrt{1-u{}_c^2} \right)~, \\
 \barP &= \frac{1}{8\pi G} 
 \left(\frac{3-u{}_c^2}{\sqrt{1-u{}_c^2}} - \frac{c{}_1}{2}\right)~, \\
 \bars &= \frac{\barepsilon+\barP}{\tilT}
 = \frac{u_c^{3/2}}{4G}~, \\
 c_s^2 &= \frac{\partial \barP}{\partial\barepsilon} 
  = \frac{1+u{}_c^2}{3(1-u{}_c^2)}~,
\end{align}
\end{subequations}
where $c_1$ is the counterterm dependence \eqref{eq:CT} ($c_1=6$ for asymptotically AdS$_5$ spacetime). In the above expressions, one can eliminate $u_c$ by proper temperature $\tilT$, but the result is not very illuminating. 

Note that the stress tensor is no longer traceless. Also, one always has $\barepsilon<0$ for $c_1=0$, which may be troublesome, but $\barepsilon>0$ for $c_1=6$. On the other hand, $\barP>0$ for both values of $c_1$. 

The computation of thermodynamic quantities has some differences in the AdS/CFT duality. The \BY tensor is the stress tensor with respect to the intrinsic metric on the surface, and it is natural to use the proper temperature. In the AdS/CFT duality, one identifies the gauge theory metric $\gamma_{\mu\nu}$ as $g_{\mu\nu} =(r/L)^2 \gamma_{\mu\nu}$. As a result, it is natural to use the Hawking temperature in the AdS/CFT duality. The field theory stress tensor is defined with respect to $\gamma_{\mu\nu}$. Then, the AdS/CFT stress tensor $\BYT_{\mu\nu}^\text{(GKPW)}$ is related to the \BY tensor as
\be
\BYT_{\mu\nu}^\text{(GKPW)} = - \frac{2}{\sqrt{-\gamma}} \frac{\delta S}{\delta \gamma^{\mu\nu}} 
\sim \left(\frac{r}{L}\right)^2  \BYT_{\mu\nu}^\text{(BY)}
%
\ee
for $p=3$. However, physical quantities from the \BY tensor in terms of the proper temperature take the same from as  the standard AdS/CFT expressions in the limit $r_c \rightarrow\infty$ (see below).

Consider two interesting limits, the $u_c \rightarrow 0$ limit and the $u_c \rightarrow 1$ limit. They correspond to the low-$\tilT$ limit and the high-$\tilT$ limit, respectively.
\begin{enumerate}

\item
In the AdS/CFT limit ($u_c \rightarrow 0$), thermodynamic quantities take the same form as the standard AdS/CFT result:
\begin{align}
\barepsilon &= \frac{3}{16\pi G} (\pi\tilT)^4~, &
\barP &= \frac{1}{16\pi G} (\pi\tilT)^4~, &
\bars &= \frac{1}{4 G} (\pi\tilT)^3~, &
c_s^2 &= \frac{1}{3}~,
%
\end{align}
where we used $c_1=6$.

\item
In the Rindler limit  ($u_c \rightarrow 1$), they reduce to Eqs.~\eqref{eq:Rindler_thermodynamic}:
\begin{align}
\barepsilon &= \frac{c_1}{16\pi G}~, &
\barP &= \frac{\tilT}{4G} - \frac{c_1}{16\pi G}~, &
\bars &= \frac{1}{4 G}~, &
c_s^2 &\rightarrow \infty~,
%
\end{align}
where we left the $c_1$-dependence for comparison with the membrane paradigm. 

\end{enumerate}

\subsection{Sound mode perturbations}

We consider sound mode perturbations in the SAdS$_5$ black hole. 
Again we take the gauge  $h{}_{* u} = 0$, and the metric is given by
\begin{align}
ds_5^2 &= \frac{1}{u}
  \left[-f (1+\htt) dt^2 + \sum_i (1+\hii) dx_i^2 + 2 \hzt dt dz \right] 
  + \frac{du^2}{4u^2f}~. 
%
\end{align}
Like the Rindler analysis, one can obtain the master equation for the master field after some algebra. The definition of the master field is not unique, but different definitions are related to each other describing the same physics. We take the following combination for the master field: 
\begin{equation}
 \Phi(u) = h{}^x{}_x + f \frac{4{h{}^x{}_x}' + 2{h{}^z{}_z}'}{4q^2 - 3f'}~,
  \label{eq:MasterField_SAdS}
\end{equation}
which obeys the following master equation: 
\begin{equation}
 \left[ u{}^{-1} f(u) \Phi(u)'\right]' + V(u) \Phi(u) = 0~, 
 \label{eq:master_SAdS}
\end{equation}
where 
\begin{align}
 V(u) &= \
\frac{ u \nw^2 \left(4\nq^2 - 3f'\right)^2 
 + \nq^2 f \left(15 u f'^2 - 36 f f'\right)
 +\nq^4 f \left(16 f - 8 u f'\right)-16 \nq{}^6 u f }{ u^3f\left(4\nq^2 - 3f'\right)^2 }~,
 \label{eq:Potential}
\end{align}
$\nw:=\omega/(2\pi T)=\omega/2$, and $\nq:=q/(2\pi T)=q/2$.

We again solve the master equation by imposing (i) the incoming wave boundary condition at the horizon $u=1$ and (ii) the Dirichlet boundary condition at $u=u_c$, $h{}^\mu{}_\nu(u_c)$. After imposing the former boundary condition, the solution takes the form
\begin{equation}
 \Phi(u) = C F(u)~. 
\end{equation}
The remaining integration constant $C$ is fixed by the boundary condition $h{}^\mu{}_\nu(u_c)$. 
We expand the solution in $\nw$ and $\nq$:
\begin{align}
F(u) &= (1-u){}^{-i\nw/2} 
   \left[F_{00}(u)
   + \left( \nw F{}_{10}(u) + \nq F_{01}(u) \right) \right. \nonumber \\
&\quad
 \left. + \left( \nw^2 F_{20}(u) + \nw \nq F_{11}(u) + \nq^2 F{}_{02}(u) \right) 
   + \cdots  \right]~.
\end{align}
Here, we factorized $ (1-u){}^{-i\nw/2}$ to implement the incoming wave boundary condition at the horizon. Then,  the incoming wave boundary condition becomes the regularity condition for $F_{ij}(u)$ at the horizon. One can easily check $F_{00}=1$. The master equation has no terms with odd powers in $q$, so one can set $F_{01}=F_{11}=0$ without loss of generality. The solutions are
\begin{subequations} 
\label{eq:Sol4Hydro}
\begin{align}
 F{}_{10} &= -\frac{i}{2}\ln(1+u)~, 
\\
 F{}_{02} &= -\frac{2}{3u} + \frac{1}{3}\ln(1+u)~, 
\\
 F{}_{20} &= 
 \frac{1}{2} \mathrm{Li}{}_2\left(\frac{u+1}{2}\right)
 - \frac{1}{2} \{\ln2-\ln (1+u) \} \ln (1-u) 
 +\ln (1+u) \left\{ \frac{1}{8} \ln (1+u)+1-\ln2 \right\}~. 
\end{align}
\end{subequations}
The integration constant $C$ is fixed by the boundary condition $h{}^\mu{}_\nu(u_c)$. Using the definition of the master field \eqref{eq:MasterField_SAdS} and the Einstein equation, we obtain $C=C_\text{num}/C_\text{den}$. [See Eqs.~\eqref{eq:ExpC} for the detailed form of $C_\text{num}$ and $C_\text{den}$.]

The response of the \BY tensor is given by
\begin{subequations}
\label{eq:BY_SAdS}
\begin{align}
\dBYtt &= - \frac{\barepsilon+\barP}{2} \frac{f}{\uc} \sumh'~,
 \\
\dBYtz &=  (\barepsilon+\barP)\left( \hzt + \frac{f}{2\uc} {\hzt}' \right)~,
 \\
\dBYxx &= - \frac{\barepsilon+\barP}{2} \frac{f}{\uc} (-{\hxx}'+{\htt}'+{\sumh}')~,
 \\
\dBYzz &= - \frac{\barepsilon+\barP}{2} \frac{f}{\uc} (-{\hzz}'+{\htt}'+{\sumh}')~,
%
\end{align}
\end{subequations}
where we used $\barepsilon+\barP = \uc^2/(4\pi G f^{1/2})$. 
In order to compare the \BY tensor with the hydrodynamic tensor, one needs to rewrite ${h{}^\mu{}_\nu}'$. Using \eq{MasterField_SAdS} together with three constraint equations, one can write ${h{}^\mu{}_\nu}'$ in terms of $h{}^\mu{}_\nu$ and $\Phi$, schematically in the form of 
\begin{equation}
 {h{}^\mu{}_\nu}' = {A{}^\mu{}_\nu}{}_\alpha{}^\beta h{}^\alpha{}_\beta 
  + B{}^\mu{}_\nu CF~. 
  \label{eq:dhs}
\end{equation}
[See Eqs.~\eqref{eq:ExpAB} for the detailed form of $A$ and $B$.]
Substituting \eq{dhs} into Eqs.~\eqref{eq:BY_SAdS}, one obtains the \BY tensor written in terms of $\hmn(\uc)$, schematically in the form of
\be
\dBYmn = \dBYmn(h^\rho_{~\sigma}(\uc), \omega, q)~.
\label{eq:BY_SAdS_full}
\ee
We omit the detailed form of the \BY tensor since they are rather cumbersome expressions. 

One can compare the \BY tensor with hydrodynamics just like the $q\neq0$ Rindler case. However, the full form of the \BY tensor is rather complicated, so we focus on the following two cases below. First, we consider the $q\rightarrow0$ limit and compare the \BY tensor with the hydrodynamic stress tensor \eqref{eq:hydro_q=0}. Second, we consider the $q\neq0$ case and extract the sound pole. 

\subsection{Homogeneous perturbations}


We first consider homogeneous perturbations. Take the $\nq\to 0$ limit in Eqs.~\eqref{eq:ExpAB} and then expand it in $\nw$. One obtains
\begin{subequations}
\label{eq:dh}
\begin{align}
{\htt}' &= 
\frac{u \left(3-u^2\right)}{3f^2} \sumh + \mathcal O\left(\nw^2\right)~, \\
{\hzt}' &= 0~, \\
{\hxx}' &= 
-\frac{u}{3f} \sumh + \frac{i\nw u (\hxx-\hzz)}{3f} + \mathcal O\left(\nw^2\right)~, 
\label{eq:dhxx}\\
{\hzz}' &= 
-\frac{u}{3f} \sumh - \frac{2i\nw u(\hxx-\hzz)}{3f} + \mathcal O\left(\nw^2\right)~. 
\end{align}
\end{subequations}
Substituting them into the \BY tensor \eqref{eq:BY_SAdS}, one obtains
\begin{subequations}
\label{eq:BY_SAdS_q=0}
\begin{align}
\dBYtt &= \frac{\barepsilon+\barP}{2} \sumh~,
 \\
\dBYtz &= (\barepsilon+\barP) \hzt~,
 \\
\dBYxx &=
 - \frac{\barepsilon+\barP}{2} c_s^2 \sumh
 + \frac{i\tilomega \uc^{3/2}}{16\pi G} \left(\hxx - \frac{1}{3} \sumh \right)~,
 \\
\dBYzz &= 
 - \frac{\barepsilon+\barP}{2} c_s^2 \sumh
 + \frac{i\tilomega \uc^{3/2}}{16\pi G} \left(\hzz - \frac{1}{3} \sumh \right)~.
%
\end{align}
\end{subequations}
Again it is not necessary to distinguish $x^\mu$ and $x^{\tilm}$ for $\dBYtt$ and $\dBYij$. Since $\dBYtz \propto \hzt$, $\dBYtz$ takes the same form in proper coordinates. 
Equations~\eqref{eq:BY_SAdS_q=0} take the same form as the hydrodynamic stress tensor \eqref{eq:hydro_q=0} with
\be
\eta = \frac{u_c^{3/2}}{16\pi G}~, \qquad
\zeta = 0~,
%
\ee
which satisfies $\eta/s=1/(4\pi)$.

\subsection{Inhomogeneous perturbations and sound pole}\label{sec:SAdS_q}

We now consider the $q\neq0$ case and the sound pole in the \BY tensor. The  coefficients of $\hmn$ and ${\hmn}'$ of the \BY tensor do not have non-trivial singularities. Thus, the pole can appear in ${\hmn}'$. But ${\hmn}'$ can be written by \eq{dhs}, so the pole can  appear in the integration constant $C$. Namely, the pole is given by $C{}_\text{den} = 0$. 
From Eqs.~\eqref{eq:ExpC}, the hydrodynamic pole is located at
\begin{equation}
 \nw = d{}_1 \nq + d{}_2 \nq^2 + d{}_3 \nq^3 + \cdots~, 
\label{eq:dispersion_SAdS}
\end{equation}
with 
\begin{subequations}
\begin{align}
 d{}_1 &= \sqrt{\frac{1+u{}_c^2}{3}}~, \\
 d{}_2 &= -i\frac{1}{3}(1-u{}_c^2)~, \\
 d{}_3 &= (1-u{}_c^2) \frac{(1+u{}_c^2)[3-2 \ln 2 + 2\ln (1+u{}_c)]-2u{}_c}
 {6\sqrt{3(1+u{}_c^2)}}~. 
\end{align}
\end{subequations}
In terms of proper quantities,
\be
\tilomega = 
\frac{d_1}{f^{1/2}} \tilq 
+ \frac{d_2}{f} \frac{\tilq^2}{2\pi \tilT} 
+ \frac{d_3}{f^{3/2}} \frac{\tilq^3}{(2\pi \tilT)^2} + \cdots~.
\label{eq:dispersion_proper}
\ee

We compare this pole with 
the dispersion relation of hydrodynamic sound mode \cite{Natsuume:2007ty}: 
\begin{align}
 \omega &= c{}_s q 
 - i \left(\frac{p-1}{p}\hateta + \frac{1}{2}\hatzeta \right) q^2
\notag\\
&\quad + \frac{1}{2 c{}_s} 
 \biggl[
 \frac{p-1}{p} \hateta 
 \left(2 c{}_s^2 \tau{}_\pi - \frac{p-1}{p}\hateta \right) 
 + \zeta
 \left(
 c{}_s^2 \tau{}_\Pi - \frac{p-1}{p} \hateta - \frac{1}{4}\hatzeta
 \right)
 \biggr] q^3
  + \cdots~. 
\label{eq:dispersion_sound}
\end{align}
The $O(q^3)$ terms are the modification by the second-order hydrodynamics. The coefficients $\tau_\pi$ and $\tau_\Pi$ are two coefficients appearing in the second-order hydrodynamics. The coefficient $\tau_\pi$ gives the relaxation time of the shear stress. 

Comparing \eq{dispersion_sound} and \eq{dispersion_proper} and using $\eta/s=1/(4\pi)$, we obtain%
\footnote{In second-order hydrodynamics, $\tau_\Pi$ is defined as $\tau_\Pi \propto \zeta$, so $\tau_\pi$ vanishes automatically.}
\begin{subequations}
\begin{align}
c_s^2 &= \frac{1+u{}_c^2}{3(1-u{}_c^2)}~, \\
\zeta &= 0~, \\
\tau_\pi &= \frac{(1+u{}_c)[1-\ln 2 + \ln(1+u{}_c)]+1-u{}_c}
 {2\pi \tilT (1+u{}_c)^2}~. 
\end{align}
\end{subequations}
The speed of sound $c_s$ agrees with the thermodynamic result \eqref{eq:SAdS_thermodynamic}. The second-order coefficient $\tau_\pi$ behaves as follows:
\be
\tau_\pi = \frac{2-\ln2}{2\pi \tilT} \quad (u_c\rightarrow0)~, \qquad
\tau_\pi = \frac{1}{2\pi \tilT} \quad (u_c\rightarrow1)~.
%
\ee
The $u_c\rightarrow0$ limit takes the same form as the standard AdS/CFT result.

\section{Discussion}\label{sec:comparison}

\subsection{Relation between Rindler and SAdS results}

In \sect{Rindler}, the Rindler result gives an incompressible fluid. On the other hand, the SAdS result gives $\zeta=0$ even in the near-horizon limit $u_c\rightarrow1$. An incompressible fluid is different from a fluid with $\zeta=0$. To answer to the question, let us study the relation between the SAdS black hole and Rindler space.

The Rindler limit of the SAdS$_5$ black hole is given by 
\begin{align}
 u &= 1 - 8 \epsilon^2 r~, &
 t &= \frac{1}{4}\, t^\NH~, &
 x_i &= \epsilon\, x_i^\NH~,&
 & \text{with } \epsilon \to 0~.
\end{align}
Note that $x_i$ is $\epsilon$-rescaled, but $t$ is not. The coefficient of $1/4$ in the definition of $t^\NH$ is necessary to match the SAdS$_5$ Hawking temperature $T_\text{SAdS} = 1/\pi$ with the Rindler temperature $T_\text{Rindler} = 1/(4\pi)$. Under the rescaling, the SAdS$_5$ metric becomes the Rindler metric up to an overall rescaling:
\be
ds^2_\text{SAdS} = \epsilon^2 ds^2_\text{Rindler}~.
%
\ee

Consider the perturbations under the rescaling. The momentum is rescaled as 
\be
\omega = 4 \omega_\NH~, \qquad 
q = \frac{1}{\epsilon} q_\NH~. 
\label{eq:scaling_q}
\ee
Since $x_i$ is rescaled but $t$ is not, $h{}^z{}_t$ must be rescaled as 
\begin{equation}
 h{}^z{}_t = 4\epsilon\, {h{}^z{}_t}{}^{\NH}~. 
\end{equation}
The other components are not $\epsilon$-rescaled since the upper and lower indices receive the opposite rescaling. 
In the Rindler limit, the SAdS master equation \eqref{eq:master_SAdS} becomes 
\begin{equation}
  \left[r \Phi' \right]' + \left(\frac{\omega_{\NH}^2}{r} - q_{\NH}^2\right)\Phi = 0~, 
\end{equation}
which is identical to the Rindler master equation \eqref{eq:master}. 
The Dirichlet boundary condition for the master field is also identical to the Rindler case. Using Eqs.~\eqref{eq:ExpC}, 
\begin{equation}
 C = \left.\frac{h{}^x{}_x}{F} + \mathcal O(\epsilon^2) \right|_{u_c}~, 
 \label{eq:BC_Rindler2}
\end{equation}
which reduces to \eq{BC_Rindler}. Thus, the SAdS master field is completely identical to the Rindler master field. 
Note that we take the $\epsilon\rightarrow0$ limit while keeping $\omega_{\NH}$ and $q_{\NH}$ fixed. Thus, $q\rightarrow\infty$ in this limit.

Even though the SAdS master field is identical to the Rindler master field, why does the SAdS ($\zeta=0$) result differ from the Rindler (incompressible) result?
In the sound mode computation in \sect{SAdS_q}, we looked at the hydrodynamic regime $\omega \sim O(q)$. But this does not mean $\omega_\NH \sim O(q_\NH)$ in the Rindler limit because of the scaling \eqref{eq:scaling_q}. 
The Rindler hydrodynamic regime does not correspond to the SAdS hydrodynamic regime. As mentioned in the last paragraph, the Rindler hydrodynamic regime actually corresponds to the short wavelength regime $q\rightarrow\infty$ in the SAdS hydrodynamic variable. 
In fact, one can show that the full SAdS \BY tensor \eqref{eq:BY_SAdS_full} reduces to the Rindler \BY tensor \eqref{eq:BY_Rindler_q} in the $\epsilon\rightarrow0$ limit. Namely, {\it the full SAdS \BY tensor contains not only the SAdS $\zeta=0$ hydrodynamics but also the Rindler incompressible hydrodynamics.}
Thus, there are no contradictions between the SAdS and Rindler results.

%
%


\subsection{Other comments}

In this paper, we give the hydrodynamic stress tensor via metric perturbations. Since we determined the velocity field via metric perturbations in \sect{hydro}, one could compare our velocity field with the fluid velocity field in the other approaches such as the BKLS approach. The BKLS approach \cite{Bredberg:2011jq} uses the ingoing Eddington-Finkelstein coordinates and does not use our gauge condition $h_{*r}=0$, but it is not difficult to make the coordinate transformation to our gauge choice. However, in the BKLS approach, the incompressible condition is $\del_\mu u^\mu=0$, whereas our incompressible condition is $\nabla_\mu u^\mu=0$ [\eq{incompressible}]. This is because the BKLS approach takes the boundary condition that the induced metric is flat. It would be interesting to extend their results for the curved induced metric in order to compare with our results.

Finally, incompressible fluids are interesting subjects to study in hydrodynamics, but one should note that the Rindler fluid is different from nonrelativistic incompressible fluids in reality. Namely, the standard nonrelativistic fluids have $\varepsilon \gg P$, whereas Rindler fluid has $\varepsilon=0$ when $c_1=0$, so it does not satisfy $\varepsilon \gg P$.

\section*{Acknowledgements}
The research of MN and TO were supported in part by a Grant-in-Aid for Scientific Research (23540326) from the Ministry of Education, Culture, Sports, Science and Technology, Japan. The research of YM and MO were supported in part by JSPS Research Fellowships for Young Scientists, No. 23-2195 and 24-5519, respectively.

\appendix

\section{Some expressions used in text}\label{sec:appendix}

\begin{itemize}

\item
The integration constant $C$ of the master field in terms of boundary values $h{}^\mu{}_\nu(u_c)$:\begin{subequations}
\label{eq:ExpC}
 \begin{align}
& C = \frac{C{}_\text{num}}{C{}_\text{den}} 
 \\
& C{}_\text{num} = 
 u (4 \nq^2 - 3 f')^2 
 \bigl[-2 \nq \nw h{}^z{}_t + \nq^2 f h{}^t{}_t
 -\{ \nq^2(1+u^2) - \nw^2 \}h{}^x{}_x - \nw^2 h{}^z{}_z \bigr] \Bigr|_{u{}_c}
 \\
& C{}_\text{den} =
  12\nq^2uf^2(4\nq^2-3f')F' \Bigr|_{u{}_c}
\nonumber \\
&\quad
 + 4[4 \nq{}^6 u (-3 + u^2) + 27 u{}^3 \nw^2 
 - 9 \nq^2 u^2 (u + u{}^3 - 4 \nw^2) 
 -  12 \nq{}^4 (1 + u{}^4 - u \nw^2)]F \Bigr|_{u{}_c}
\end{align}
\end{subequations}

\item
Explicit expressions of \eq{dhs}:
\begin{subequations}
\label{eq:ExpAB}
\begin{align}
 {h{}^t{}_t}' 
 &= \frac{1}{6f^2}
 \Bigl[
 - 4 \nq^2 h{}^t{}_t
 + [8 \nw^2 + 4 \nq^2 (f - uf') - 3 (3 f-u f')f']h{}^x{}_x
\nonumber \\
&\quad
 + 4 \nw^2 h{}^z{}_z
 + 8 \nq \nw h{}^z{}_t
 - (4 \nq^2 - 3 f') (3 f - u f') CF
 \Bigr]~,
\\ 
 {h{}^x{}_x}' 
 &= \frac{1}{12\nq^2 f^2}
 \Bigl[
 \nq^2 f (4 \nq^2 - 3 f') h{}^t{}_t
 + [(4 \nq^2 - 3 f')(\nw^2 + \nq^2 u f') -(4 \nq^4 - 9 \nq^2 f')f ] h{}^x{}_x
\nonumber \\
&\quad
 - \nw^2 (4 \nq^2 - 3 f') h{}^z{}_z
 - 2 \nq\nw(4 \nq\nw - 3 f')h{}^z{}_t
\nonumber \\
&\quad
 - ( 4 \nq^2 -3 f')(3 \nw^2 - 3 \nq^2 f + \nq^2 u f') CF
 \Bigr]~,
\displaybreak 
\\
 {h{}^z{}_z}' 
 &= \frac{1}{6 \nq^2 f^2}
 \Bigl[
 - \nq^2 f (4\nq^2 - 3 f') h{}^t{}_t
 - [8 \nq^4 f + (4 \nq^2 - 3 f')(\nw^2 + \nq^2 u f')]h{}^x{}_x
\nonumber \\
&\quad
 + \nw^2 ( 4 \nq^2 - 3 f') h{}^z{}_z
 + 2 \nq\nw (4 \nq^2 - 3 f') h{}^z{}_t
\nonumber \\
&\quad
 + (4 \nq^2 - 3 f')(3 \nw^2 + \nq^2 u f') CF
 \Bigr]~,
\\ 
 {h{}^z{}_t}' 
 &= \frac{1}{2 \nq f}
 \Bigl[
 \nw ( 4 \nq^2 -f')h{}^x{}_x
 + \nw f' h{}^z{}_z
 + 2 \nq f' h{}^z{}_t
 - \nw (4 \nq^2 - 3 f') CF
 \biggr]~.
\end{align}
\end{subequations}

\end{itemize}


\end{document}